# Reversing the asymmetry in data exfiltration

David Skillicorn, Xiao Li, Karen Chen
School of Computing, Queen's University

Abstract: Preventing data exfiltration from computer systems typically depends on perimeter defences, but these are becoming increasingly fragile. Instead we suggest an approach in which each at-risk document is supplemented by many fake versions of itself. An attacker must either exfiltrate all of them; or try to discover which is the real one while operating within the penetrated system, and both are difficult. Creating and maintaining many fakes is relatively inexpensive, so the advantage that typically accrues to an attacker now lies with the defender. We show that algorithmically generated fake documents are reasonably difficult to detect using algorithmic analytics.

## Exfiltration

Organisations often hold data of importance to the enterprise online, and such data can be the target of state actors, competing businesses, and cybercriminals. Exfiltration – the illicit removal of data from within the organisation's control – is a serious risk for many organisations. As in many cybersecurity settings, attackers have an inherent advantage – they only have to find one loophole for exfiltrating the data, while the organisation has to defend against all possible loopholes.

Almost all defensive approaches rely on some kind of technology embedded in a perimeter. For example, firewalls may be configured to block traffic to unknown IP addresses, or to block unexpected large volumes of data transfer. Document management tools can be configured to insert (invisibly) special codes in protected documents, and the firewall configured to block transfers of documents containing these codes. However, once an attacker has gained access to the system, it is difficult to prevent exfiltration using, for example, low and slow techniques such as concealing data inside apparently innocuous web or DNS traffic (Born, 2010, Haycraft 2017, Silowash et al. 2013). Exfiltration that uses non-network mechanisms, such as high-frequency audio signals, is also hard to defend against. Ullah et al. (2017) provide a recent survey. It is also relatively straightforward to exfiltrate data by copying it onto a removable storage device, which might be only the size of a fingernail – typical cybersecurity approaches do not help prevent this, one reason why IBM has banned removable storage from their systems entirely.

We propose, and demonstrate, a different approach to the problem. Rather than trying to prevent a particular important document from being exfiltrated, we instead create multiple 'fake' versions of the document. The creation process is automated, so that it is possible to generate many thousands of 'fakes' at low cost. The intention is to make the fakes good enough that it is difficult to distinguish them from the real document algorithmically. Humans may still be able to detect the real among the fakes, but this requires exfiltrating and reading all of them. As the number of fakes increases, this becomes more and more difficult. We do not try to make the fakes more attractive or more believable or more conspicuous, that is they are not honeypots or decoys. We also assume an attacker, whether from outside or an insider, has enough privilege that they can see everything done by the system (unlike Bowen et al., 2009).

Decoys have been previously used in cybersecurity contexts, but their primary purpose has been to act as traps – decoys are instrumented to detect unauthorized access, and perhaps exfiltration, because there is no legitimate reason why they should be accessed (Voris et al. 2013). Honeypots have also often been used; but the purpose of a honeypot is to divert an attacker to different, apparently higher-value content and/or use the honeypot as a decoy.

Our approach is different – we create fakes, not to detect attempts to access them, but to make the wheat expensive to find amongst the chaff. This reverses the asymmetry that is typical of adversarial settings – the defender can create arbitrarily large amounts of work for the attacker. Of course, there are costs to this approach: the fakes require extra storage (but this is relatively cheap); and computation has to be spent managing the fakes (but this is small).

Legitimate users who interact with documents must also be able to distinguish the real version among the fakes. We suggest that this be done using secret sharing, so that the system itself does not know which version is the real one (until a user asks for it, and provides the other part of the secret). To prevent traces of user interaction becoming detectable, the system also permutes the files after each save operation, so that the name of the real document constantly changes.

There are many potential approaches to creating fake documents. To demonstrate that the approach is viable, we used a red team/blue team approach. One of the authors built a system for creating and managing fake versions (as well as the secret sharing required to identify the real one) while another built a system that tries to detect the real among the fakes. This practical experiment provides some hints about the best techniques for creating effective fakes. Of course, there is conceptually an arms race between fake-building and fake-detecting algorithms, but our primary purpose is to show that it is possible to build fakes that are reasonably difficult to detect.

# Fake creation techniques

We assume that an attacker has some idea of the content of the targeted document, and so may first try to filter out documents that do not contain a reasonable number of expected key words. Thus it is not a viable strategy simply to create random documents. The fake documents must be built from the real document that they are mimicking. Some experimentation suggested that fake documents should overlap real ones in about 70% of the words present.

The kinds of key information contained in real-world documents is seldom limited to a few specific words ('the formula' that appears in many spy narratives); rather it is strategic or tactical content that is embodied in the language in many different places. Even in a document such as a tender, where the bid price is a critical piece of information, a context of terms and conditions is needed to put the price in context. We focus on creating fakes using changes throughout the target document, rather than trying to find critical regions.

## Paragraph and sentence substitutions

For each type of document, a user can provide sample paragraphs and sentences, forming a bank from which insertion or substitutions can be made. The content of this bank does not matter much, since word substitutions happen later, but it can be used to insert particularly meaningful or misleading content that is, of course, only included in the fakes. Each paragraph and sentence of the original document are considered for replacement or expansion based on the overlap factor.

## Single word substitutions

The overlap factor determines the probability that each noun will be replaced. Changes to the document cannot be simply, say, random word replacements because these create discontinuities that are much too easy to detect. The pair and triple frequencies in English have been computed; a random replacement is likely to create a set of low frequency 2- and 3-grams that are suspicious. We try to find replacements that are not obviously out of place, at least from an automated analysis perspective.

There are three possible replacement strategies: replace by synonym, replace by antonym, or replace by associated word. These replacements muddle the content while still seeming statistically appropriate.

The system maintains a cache of possible replacements, since it will encounter the same words in creating each fresh fake. The first time a word is considered for replacement, its synonyms and antonyms are collected using the Wordnet API, and associated words using the Words API. All of these words are collected into a JSON object, and a random element selected as the actual replacement. If the word has been encountered previously, then only the last step is necessary – a random element from the existing JSON object is selected.

After substitution, a grammar checking API is used to ensure that the replacement is grammatical. For example, some of the possible replacements for 'addition' are: 'decrease', 'lessening', 'loss', 'reduction', 'shrinkage', 'extension', 'inclusion', 'enlargement', 'annexation', 'augmentation', and 'moreover', and all but the last are plausible.

Here is an example of a paragraph from the real document, and one of its fakes:

Original: Current theories focus on personal characteristics to explain wrong-doing and how someone can intentionally harm others. In a survey, professionals such as doctors, psychologist and laymen predicted that a small proportion of a population (1-3%) would harm others if ordered to do so.

Fake: Freshly supposition focused on exclusive tendency to explain wrong doing and how star can intentionally harm contrasting. Modern an ignorance, professionals such as scientist, psychologist and laymen foreshadow that a small extent of a census (1 3) would damage lack if ordered to do extremely.

Clearly the fake version is gibberish to a human reader, but it is quite difficult to detect as gibberish using automated analysis.

Benford's law and dates

It is well known that the digit distribution for real-world numbers has a typical shape. In particular, the probability that the first digit of such a number is 1 is close to 30%, a 2 close to 15% and so on, with the probability of an initial 9 only 4.5%. There are similar, weaker expected distributions for second and subsequent digits. This property is known as Benford's Law (Benford, 1938). It is regularly used by tax departments, customs, and auditors to detect potentially made up numbers.

Benford's law means that fake numbers cannot simply be randomly generated if they are to seem plausible. Instead, we generate the leading four digits of each fake number using the Benford's Law probability distribution.

In some exfiltration settings, it is numeric values such as credit card numbers that are the high value content, so there is a particular attraction to altering them. Credit card numbers contain a check digit, so they should in fact be treated specially; but uniformly random changes are probably undetectable enough by most cybercriminals that they remove any value from the credit card numbers.

Dates are a special form of numeric data that must obey particular rules – about possible times of day, and possible days of the month. The system changes dates in documents by adding or subtracting random values to days, months and years. However, to prevent rediscovery of the real date by taking the midpoint of the distribution of the altered dates across documents, the system uses a biased change mechanism that makes dates likely, on average, to be earlier than the real value from which they are adapted.

Time stamp management

If the fakes are based on the real document then, without some care, all of the fakes would have timestamps later than the real document, making the real document easy to detect.

This problem is addressed first by giving each file, including the real one, a random system name using characters and digits. Then all of the files are modified in trivial ways using the order based on their random names, and saved. Thus the time stamps depend only on the random name given to each version. (Users access files using the secret sharing mechanism and so do not need to know the actual file name.)

A similar problem happens when a user legitimately edits the real document. In order to do so, they must use the secret sharing interface (described below) so there is an opportunity for the system to detect a potential change to the real document. In the background, it makes trivial changes to the fake versions with random intervals between each, and ensuring that some happen after the user's final save of the real document. The fakes only need to be manipulated in a time window around actions by a user on the real document; a background process is not required.

Thus at no time is the timestamp of the real document consistently different from those of the fakes.

Secret sharing
Since we assume that an attacker has access to the host system, the identification of which is the real file and which are the fakes cannot exist on the system. Instead, we use secret sharing – a secret consists of two distinct parts, both of which must be simultaneously present for authentication, one of which is kept by the system (where it is of no use by itself), and the other is kept by the user.

There are many secret sharing schemes; we use one developed by Shamir. We want a secret with two pieces, one for the system and one for the user. We first choose a line in the plane, and then give each participant the coordinates of a point on the line. Together these two points define a line, and the x-intercept of this line is the secret. Knowing either of the points reveals nothing; only knowing both at once reveals the line and hence the secret. (In fact, the calculations are done in a finite field of chosen prime order, but the intuition is the same.)

There is a 1-1 mapping of the secret to the file name. Any other value for the secret will point to some other file. The secret can be recalculated whenever the file is used; or when the file name changes. Note that there is no security issue with the user writing down their part of the secret since, by itself it is useless (unless the attacker has both physical and cyber access to the organisation). Also the user part of the secret does not need to change even when the file name changes, since the system part of the secret can be changed to match.

# Fake detection techniques

The strategy for detecting the fake versions of documents is to apply a variety of tests, each looking for a different signature of fake or genuineness, and each generating a suspicion score. These suspicion scores can then be combined to create an overall ranking from most to least suspicious. The real document should be at the top of this ranking; but it would still be useful to an attacker if the ranking tended to place the real document quite high, since only a small fraction of documents would need to be exfiltrated to include the real one.

n-gram frequencies
Pairs of words are related to one another because of the process of selectional restriction – the meaning of a given word restricts words that can appear adjacent to it to one whose meaning is somehow consistent. For example, a verb of eating (more or less) requires an object that is capable of being eaten. Thus altering a document by replacing only single random words creates pairs of words that do not fit together, and so create a detectable disfluency.

Large tables of n-grams (sets of n adjacent words) and their frequencies are readily available, although we used the Phrasefinder API to avoid storing such large tables. We looked up the expected frequencies of pairs of adverb-verbs and adjective-nouns, since these are the most likely places where substitutions create disfluencies. Low frequency pairs, especially overlapping low frequency pairs (X-Y and Y-Z) are potential signals of single word substitution.

A suspicion score for two adjacent words w1 and w2 is computed as

$$f(w1) \times f(w2) / (f(w1\ w2) + 1)$$

so that words that occur reasonably frequently individually, but not usually together, have high suspicion scores.

### Rates of new words

The rate at which new words are encountered in a document (new in the sense of not encountered so far) is a quite predictable function of the total number of words encountered so far in most documents. For example, most writing introduces a theme and then develops it, and this naturally causes the rate of new words to slow as the same words are mentioned in developing contexts. The shape of the resulting function is quite well approximated by the digamma function.

Strategies that replace large pieces of text, for example at the paragraph level, produce detectable changes in the shape of such a function. This seems to be for two reasons. The first, and obvious one, is that the topic of a replacement paragraph is unlikely to be the same as the topics of the surrounding text into which it was inserted. Therefore the nouns, adjectives, verbs, and adverbs are likely to be (locally) new, and so cause an upward turn during the inserted paragraph, perhaps followed by a flatter region after its end.

Authorship detection has shown that different authors tend to use different (patterns of) function words, those usually short but frequent words such as articles and pronouns that make up almost 50% of words used. The second reason, therefore, is that each individual author's use of function words differs. A replacement paragraph written by a different author is likely to contain a sudden tranche of new function words.

These changes are reasonably visible to the human eye in a plot of unique words encountered as a function of words encountered – a small bump appears on the plotted line. However, we were unable to develop an algorithmic technique to find such bumps. A number of measures were tried, for example looking for local increases then decreases in slope, but the changes were too subtle for these to work reliably.

The figure shows the curve for a normal document on the left, and a document with a substituted paragraph (around 300 words in) on the right.

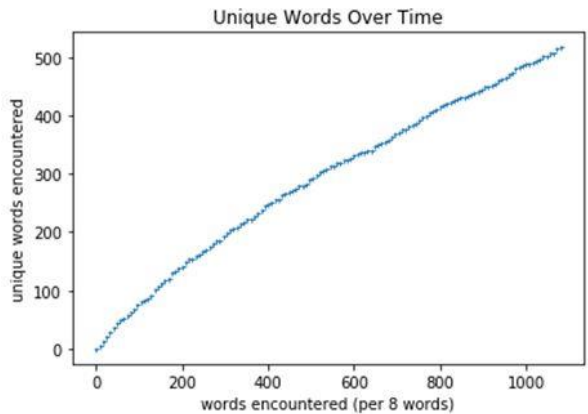 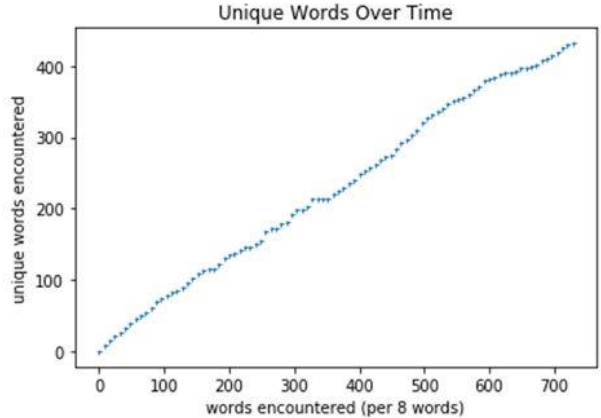

Zipf's law

Zipf's law describes the frequency histogram of words in a document when sorted from most to least frequent. Roughly speaking, if the frequency of the most frequent word is f then the frequency of the second most frequent word in a document will be proportional to 1/2 f, the frequency of the third most frequent word proportional to 1/3 f, and so on. Thus plotting words in frequency order creates a typical shape with a strong initial drop which then flattens out.

Paragraph (and to a lesser extent sentence) substitutions introduce completely new words and so create a different frequency distribution (depending, of course, on how different the replacement paragraph is). This can be partly because the content of the replacement paragraph is different – it uses different nouns and adjectives. However, this would not necessarily create a strong signal, since nouns and adjectives are not usually high frequency words.

Different patterns of function word use appear to be what generates detectable signal from paragraph replacement – it's not so much that the replacement paragraph is about a different topic as that it was written by a different person, and so the frequency histogram of the altered document differs in its use of frequent function words.

The figure shows a Zipfian distribution (typical of normal documents) overlaid with the distribution for a document containing substitutions.

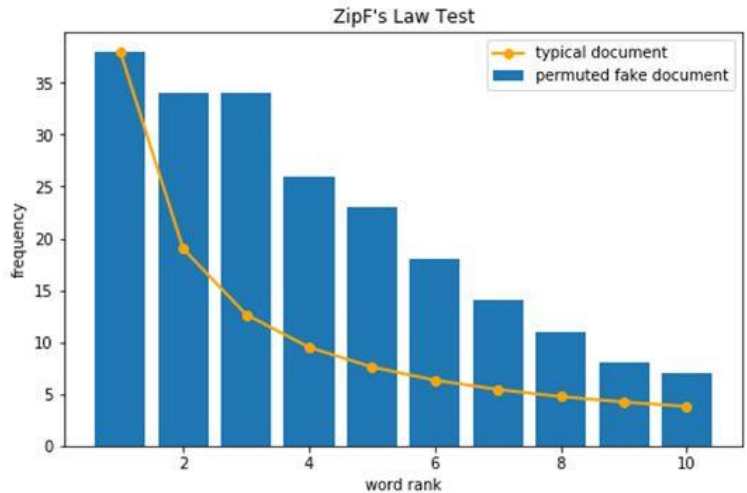

Benford's law

While Benford's Law makes strong predictions about the distribution of digit values in real-world numbers, it is hard to use this as a detector of anomalies because the number of digit strings in most documents is relatively small. This is especially true if dates are treated as a special case, as they should be.

We experimented only with examining the first digit of each non-date number. For a document, the expected Benford's Law frequencies are summed over all of the number occurrences, based on the actual digits that appear, and divided by the number of digit strings. The larger this number, the more likely that there have not been substitutions and so the less suspicious that document is. A document in which every number string begins with 1 gets a score of 0.3 (the maximum possible) while a document in which every number begins with 9 gets a score of 0.046, and so seems quite suspicious.

In practice the signal from exploiting Benford's Law is weak, but fake documents where numbers have been generated in some other way tend to have some level of suspicion associated with them.

Time stamp histograms

The real version of each document presumably changes over time and so its time stamp will also changes to reflect this. The fakes must also be manipulated so that their time stamps continue to change as well. There are many ways in which this could be done, but the general strategy must surely be to prevent the real document from looking unusual among the fakes.

In our experiments, all versions of the same document had time stamps that were identical within the resolution that the operating system made available. This does not quite mean that time stamps are useless as a detection technique – touching all of the files in a set takes longer as the size of the set increases, so there may be potential here for practically large document sets.

Singular value decomposition

The bag of words model is often used to represent documents. A corpus is described by a matrix with one row for each document and one column for each word that appears in any of the documents. The ij[th] entry of this matrix gives the frequency of word j in document i. The bag of words representation ignores word order but, despite this, produces usable representations even for languages like English where word order is important.

Some normalizations of the document-word matrix are necessary. Long documents contain more words than short ones, so the row sum of a long document's row will be greater than that of a short document. It is usual to divide the entries in each row by the sum of counts in that row, effectively converting word counts into word rates. Rates are properly comparable among documents. Let A be a normalized document-word matrix describing a set of documents, one real and many fakes.

The singular value decomposition (SVD) of a matrix, A, expresses it as a product of three matrices

$$A = U\ S\ V'$$

If A is $n \times m$, then U is an $n \times m$ orthogonal matrix, S is a diagonal matrix (i.e. off diagonal entries are zeroes), and V is an $m \times m$ orthogonal matrix. The superscript on V indicates matrix transposition. A truncated SVD provides an approximation for A as a product of an $n \times k$ U, a $k \times k$ S, and an $m \times k$ V. Each row of U corresponds to a row of A, and each row of V to a column of A. If k is chosen to be 3, then the rows of U (resp. V) can be regarded as coordinates in 3-dimensional space and a point plotted for each document.

The SVD projects the document-word matrix into a k-dimensional space in a way that reveals that greatest variation among the set of documents. A really clever manipulation of the original documents might make all of the fakes seem different both from it, and from one another. In this case, the real document might appear at the centre of a cloud of the fakes. However, this generality of manipulation is difficult, and potentially expensive; using a set of manipulations that is basically similar, but with different parameters, tends to make the real document unusual among the fakes. After the SVD embedding, therefore, a real document tends to be plotted far from the cloud of fakes.

This is illustrated in the two SVD embeddings below. In the left-hand figure, the document far from the others is indeed the real one; in the right-hand figure, the real one is labelled pLudqZZw.txt, fairly far from the centre but not the most extremal. Thus distance from the centre of the cluster of documents is suspicious but, by itself, not definitive.

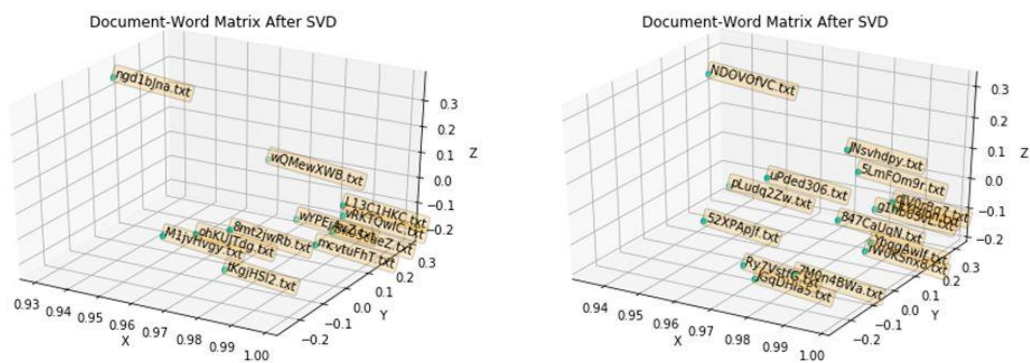

## Discussion and conclusions
This design and these experiments have shown that:

- It is possible to generate fake versions of a document with reasonable amounts of computational effort, and that the real document can be tracked among the fakes using secret sharing, while at the same time making sure that the system itself does not know which is the real document except when it absolutely must – when a user is interacting with it.
- Given the full set of documents (real + fakes) it is difficult and expensive to try and determine which is real and which are fake; and this can only be done by ranking the set by suspiciousness. If the set is too large to exfiltrate all members (which the defender can ensure) then the algorithmic work to detect the real among the fakes must be done *within* the penetrated system, itself potentially generating signal associated with the intrusion.

The costs for creating and managing the fakes are moderate – some computation to create them, manage their timestamps, process the secret, and some increased storage – but the approach defends against any exfiltration technique of any sophistication, including the use of removable storage devices.